\documentclass[twocolumn, traditabstract]{aa}
\usepackage{color}
\usepackage{soul}
\usepackage{graphicx}
\usepackage{txfonts}
\usepackage{lscape}
\usepackage{longtable}
\usepackage{threeparttable}
\usepackage{setspace}
\usepackage{caption}
\usepackage{natbib}
\bibpunct{(}{)}{;}{a}{}{,}
\sethlcolor{yellow}

\def\farcs{\hbox{$^{\prime\prime}\!\!\!.\,$}}

\begin{document}
   \title{Starbursts and black hole masses in X-shaped radio galaxies:\\
Signatures of a merger event?}
	
   \author{M. Mezcua\inst{1,2,3}
	\thanks{Email:mmezcua@iac.es}
   \and V.H. Chavushyan\inst{4}
   \and A.P. Lobanov\inst{1}\thanks{Visiting Scientist, University of Hamburg / Deutsches Elektronen Synchrotron Forschungszentrum.}
   \and J. Le\'on-Tavares\inst{5}}
   \institute{Max Planck Institute for Radio Astronomy,
              Auf dem H\"ugel 69, 53121 Bonn
   \and Instituto de Astrof\'isica de Canarias, V\'ia L\'actea S/N, 38200 La Laguna, Tenerife, Spain
   \and Dept. Astrof'sica, Universidad de La Laguna (ULL), 38206 La Laguna, Tenerife, Spain
   \and Instituto Nacional de Astrof\'isica, \'Optica y Electr\'onica, Apdo. Postal 51, 72000 Puebla, M\'exico
   \and Aalto University Mets\"ahovi Radio Observatory, Mets\"ahovintie 114, FIN-02540 Kylm\"al\"a, Finland}

  \abstract{We present new spectroscopic identifications of 12
X-shaped radio galaxies and use the spectral data to derive starburst
histories and masses of the nuclear supermassive black holes in these
galaxies. The observations were done with the 2.1-m telescope of the
Observatorio Astron\'omico Nacional at San Pedro M\'artir,
M\'exico. The new spectroscopic results extend the sample of X-shaped
radio galaxies studied with optical spectroscopy. We show that the
combined sample of the X-shaped radio galaxies has statistically
higher black-hole masses and older episodes of star formation
than a control sample of canonical double-lobed radio sources with
similar redshifts and luminosities. The data reveal enhanced star-formation activity in the X-shaped sample on the timescales expected in
galactic mergers. We discuss the results obtained in the framework of
the merger scenario.}


\maketitle

\section{Introduction}

A significant fraction of the most powerful FRII (Fanaroff $\&$ Riley
type II; \citealt{1974MNRAS.167P..31F}) radio galaxies exhibit two
pairs of misaligned radio lobes associated with the same parent
galaxy (\citealt{1984MNRAS.210..929L}), giving these galaxies a
peculiar X-shaped appearance, which is remarkably different from the
``canonical'' double-lobed morphology of radio galaxies. Hence, these
objects are often branded `winged' or `X-shaped' radio galaxies
(\citealt{1984MNRAS.210..929L}; \citealt{1992ersf.meet..307L}).  A
recent merger of two supermassive black holes (SMBHs), where the wings constitute the relic emission of the past radio jets, is one of 
several scenarios proposed to explain this peculiar radio morphology
(e.g., \citealt{rottmann}; \citealt{2002Sci...297.1310M};
\citealt{2006MmSAI..77..733K}; \citealt{2009ApJ...697.1621G};
\citealt{2010ApJ...717L..37H}; \citealt{2011A&A...527A..38M}).
Alternative models employ jet-axis reorientation (e.g.,
\citealt{2002MNRAS.330..609D}), two unresolved active galactic nuclei (AGN; \citealt{2007MNRAS.374.1085L}), backflow of material from the main lobes into the wings (e.g.,
\citealt{1984MNRAS.210..929L}; \citealt{2002A&A...394...39C};
\citealt{2010ApJ...710.1205H}; \citealt{2010MNRAS.408.1103L};
\citealt{2011ApJ...733...58H}), or non-ballistic precession assuming a pre-merger state of the two black
holes (\citealt{2008MNRAS.389..315G}; \citealt{2011ApJ...734L..32G}). The existence of very few X-shaped sources with FRI edge-darkened radio lobes (\citealt{2008arXiv0806.3518S}; \citealt{2009ApJ...695..156S}) was initially used as a strong argument against the backflow scenario. However, the presence in most of the FRI X-shaped galaxies of double-double morphologies indicative of a renewal of jet activity, together with an alignment between the main radio axis and the major axis of the host galaxy, led to the proposal of a common formation mechanism for FRI and FRII X-shaped galaxies via backflows (\citealt{2009ApJ...695..156S}). The required duration of these backflows is, however, too close to the typical radio-source lifetime of 10$^{8}$ yr (e.g., \citealt{rottmann}; \citealt{2007MNRAS.374.1085L}), which again argues against the backflow scenario.

Each of the models mentioned above thus has difficulties in explaining the entire range of the observed
properties of the X-shaped radio galaxies (see \citealt{2010arXiv1008.0789G} for a review). 
If the observed X-shaped morphology results from a coalescence of a pair
of SMBHs, the imprints of the preceding galactic merger should be
observed in the X-shaped radio sources in the form of early-type host
galaxies (\citealt{1977egsp.conf..401T}), statistically higher black
hole (BH) masses, and enhanced star-formation and nuclear activity
(e.g., \citealt{2000MNRAS.311..576K}; \citealt{2008ApJS..175..356H}).
It would therefore be relevant to investigate whether BH masses and star-formation activities differ 
between X-shaped radio galaxies and canonical double-lobed radio sources. We initiated such a study by analyzing the optical spectra of 29 X-shaped sources and 36 radio-loud control sources (cf., \citealt{2011A&A...527A..38M}) . The results revealed that all the X-shaped sources analyzed are hosted by early-type galaxies, and that the X-shaped sample has a statistically higher mean BH mass and older bursts of star formation than the control sample of canonical radio galaxies (\citealt{2011A&A...527A..38M}; hereafter M11). 
A total of 52 X-shaped radio galaxies have been spectroscopically identified so far in several studies (\citealt{2009ApJS..181..548C};
\citealt{2010MNRAS.408.1103L}), but not all of these observations are suitable for detailed spectral analysis. 
To increase the statistical significance of our studies, we identified a sample of 13 X-shaped radio sources with bright optical
nuclei suitable for optical spectroscopy observations. We present here the results of the spectroscopic observations and discuss them in the
context of physical models proposed to explain the nature of the X-shaped radio galaxies. 

The objects selected and the optical observations of the sample are
described in Sect.~2. Procedures applied for the analysis of the measurements are explained in Sect.~3. The results and
discussion are presented in Sects.~4 and ~5, respectively.
Throughout the paper, we assume a $\Lambda$ CDM
cosmology with parameters $\mathrm{H_{0}=73\ km\ s^{-1}\ Mpc^{-1}}$,
$\Omega_{\Lambda}=0.73$, and $\Omega_\mathrm{m}=0.27$.

\section{Observations and data reduction}

The observed sample comprises 13 X-shaped radio galaxies retrieved
from a list of 100 candidates (\citealt{2007AJ....133.2097C}). We
selected all the objects with bright (m$_\mathrm{R}$ $<$ 19), optical counterparts for
which neither spectroscopic redshifts had been reported nor spectra were available in the Sloan Digital Sky Survey (SDSS), and which were observable during the period of the observations (i.e., across a range of right ascension between 07:00:00 and 17:00:00). 

Optical spectroscopy observations of the sample were carried out on
20-23 March 2009 with the 2.1-meter telescope of the Observatorio
Astron\'omico Nacional at San Pedro M\'artir (OAN-SMP), Baja
California, M\'exico. The Boller $\&$ Chivens spectrograph was tuned
to the 4000 $\AA$ to 8000 $\AA$ range (grating 300 l/mm), with a spectral dispersion
of 4.0 $\pm$ 0.3 $\AA$/pix, corresponding to 8.0 $\pm$ 1.2 $\AA$ full width at half maximum (FWHM), derived from the FWHM of different emission lines of the arc-lamp spectrum. The instrumental resolution estimated for each object from the FWHM of the night-sky line at 5575 $\AA$ is in the range 9-14 $\AA$. A 2{\farcs}5 slit oriented along right ascension was
used. To calibrate the spectral measurements, the spectrophotometric
standard star Feige 34 was observed two times during every night of
the observing run. The airmass of the observations ranged between 1.0 and 1.3.

For each target source, two or three exposures were taken, with a typical
exposure duration of 1800 s (see Table~1). The data
reduction was carried out with the IRAF\footnote{IRAF is distributed by
the National Optical Astronomy Observatories operated by the
Association of Universities for Research in Astronomy, Inc. under
cooperative agreement with the National Science Foundation.} software
following standard procedures.  The spectra were bias-subtracted and
corrected with dome flat-field frames. Cosmic rays were removed interactively from
all images. Arc-lamp (CuHeNeAr) exposures
were used for the wavelength calibration. A spline function was fitted to determine the dispersion function (wavelength-to-pixel correspondence).
Sky emission lines located at known wavelengths were removed during the calibration in wavelength. The absolute flux-calibration accuracy of the spectra, provided by the standard star calibration, is about 15\%.
The relative flux-calibration uncertainty, estimated as the difference between the normalized true and estimated flux-calibration curves of the standard star, is $<15\%$ for all nights.


\begin{table}
\label{table1}
\begin{minipage}{\columnwidth}
\caption{Observation log}
\centering
\begin{tabular}{llll}
\hline
\hline 
Name & m$_\mathrm{R}$ & Obs. date & Exp. time \\
(1)  & (2) & (3) & (4) \\
\hline
J0813+4347 & 16.1 & 2009 Mar 20 & 2 $\times$ 1800 \\
	   &      & 2009 Mar 22 & 2 $\times$ 1800 \\
J0838+3253 & 16.9 & 2009 Mar 21 & 2 $\times$ 1800 \\
	   &      & 2009 Mar 23 & 2 $\times$ 1800 \\
J0924+4233 & 17.8 & 2009 Mar 23 & 2 $\times$ 1800 \\
J1008+0030 & 15.8 & 2009 Mar 20 & 2 $\times$ 1800 \\
	   &      & 2009 Mar 22 & 2 $\times$ 1800 \\
J1055-0707 & 17.6 & 2009 Mar 20 & 3 $\times$ 1800 \\
J1200+6105 & 18.3 & 2009 Mar 23 & 2 $\times$ 1800 \\
J1201-0703 & 16.4 & 2009 Mar 22 & 2 $\times$ 1800 \\
	   &      & 2009 Mar 23 & 2 $\times$ 1800 \\
J1258+3227$^{a}$ & 17.0 & 2009 Mar 20 & 2 $\times$ 1800 \\
J1351+5559 & 15.1 & 2009 Mar 22 & 2 $\times$ 1800 \\
	   &      & 2009 Mar 23 & 1 $\times$ 1800 \\
J1408+0225 & 18.4 & 2009 Mar 22 & 3 $\times$ 1800 \\
J1459+2903 & 16.5 & 2009 Mar 21 & 2 $\times$ 1800 \\
J1537+2648 & 18.3 & 2009 Mar 22 & 2 $\times$ 1800 \\
	   &      & 2009 Mar 23 & 2 $\times$ 1800 \\
J1606+0000 & 15.0 & 2009 Mar 21 & 2 $\times$ 1800 \\
\hline
\end{tabular}
\end{minipage}
\smallskip\newline {\bf Column designation:}~(1) -- object name based on J2000.0 coordinates; (2) -- apparent magnitude R, taken from \cite{2007AJ....133.2097C}; (3) observation date; (4) exposure time, in seconds. {\bf Notes:}~$a$ -- Optical misidentification in \cite{2009ApJS..181..548C}. The spectrum of this object corresponds to a K-type star and not a X-shaped radio galaxy.
\end{table}


\section{Analysis of the spectra}
The optical spectra were used to identify the most prominent emission/absorption lines, determine the galaxy redshifts, obtain mass
estimates for the nuclear SMBH, and
reconstruct histories of star-formation activity in the host
galaxies.

Since most of the observed spectra of the target objects do not show
strong emission lines, we determined the redshift from the Ca II H
$\lambda$3968$\AA$ and K $\lambda$3934$\AA$, G band
$\lambda$4302$\AA$, Mg Ib $\lambda$5175$\AA$, and Na Id
$\lambda$5896$\AA$ absorption lines.

\begin{figure}
 \centering \includegraphics[width=\columnwidth,
  clip=true]{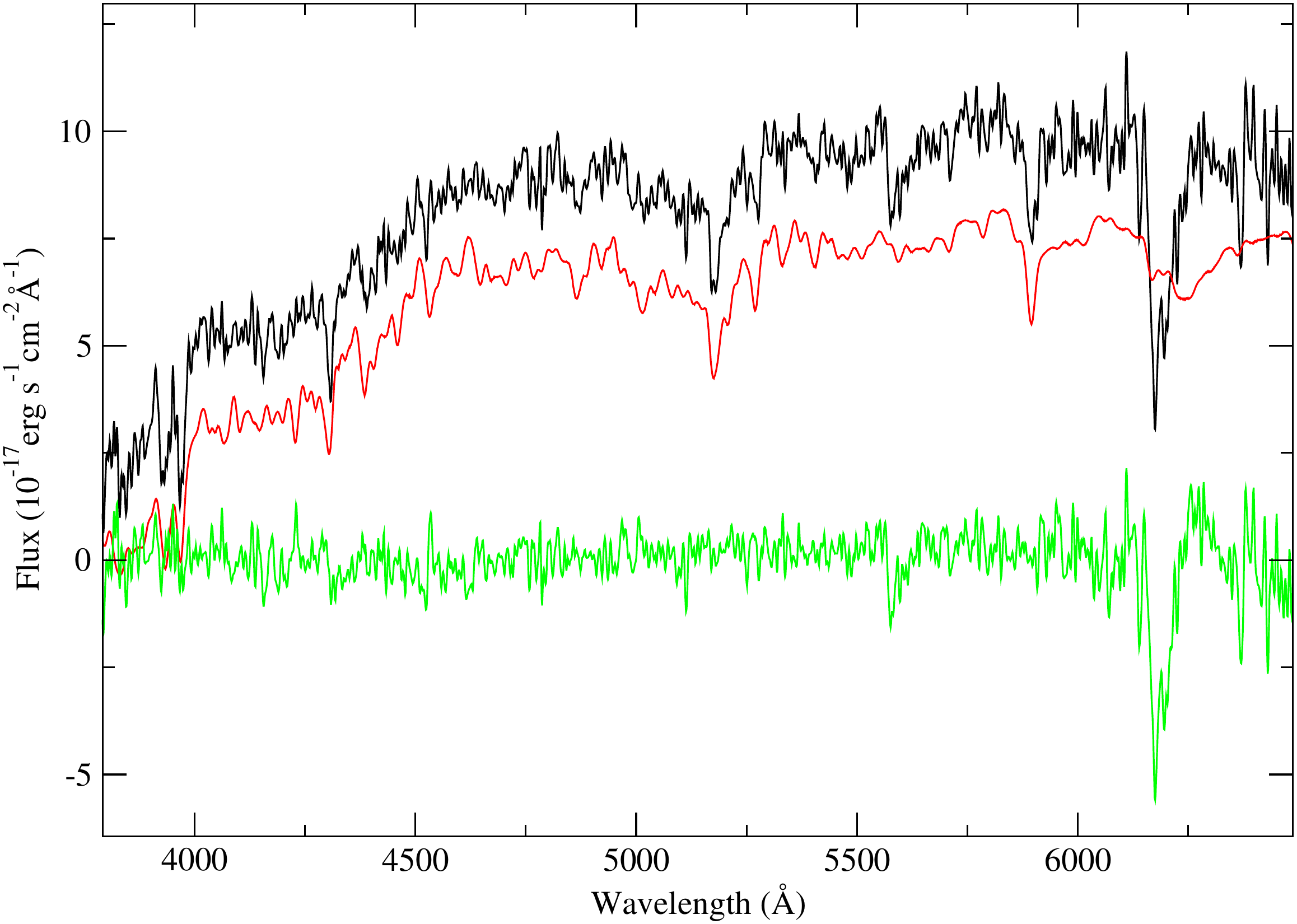} \caption{STARLIGHT fit to the
  spectrum of the X-shaped radio galaxy J1201-0703. The observed
  spectrum is shown in black, the modeled in red (for illustration
  purposes, displaced from the observed spectrum). Residuals obtained
  after subtraction of the modeled spectrum from the observed one are
  shown in green.}
\label{figure1}
\end{figure}

We used the stellar population synthesis code STARLIGHT
(\citealt{2007MNRAS.381..263A}; \citealt{2004MNRAS.355..273C,
2005MNRAS.358..363C,2007MNRAS.375L..16C};
\citealt{2006MNRAS.370..721M}) to model the observed spectra.  The
best fit was obtained by constructing a linear combination of simple
stellar populations (SSPs) from the stellar library of
\cite{2003MNRAS.344.1000B} and adding a power-law component
representing the AGN continuum emission (see Fig.~\ref{figure1}; e.g., \citealt{2011MNRAS.411.1127L}).

\begin{figure}
 \centering
  \includegraphics*[width=\columnwidth, clip=true]{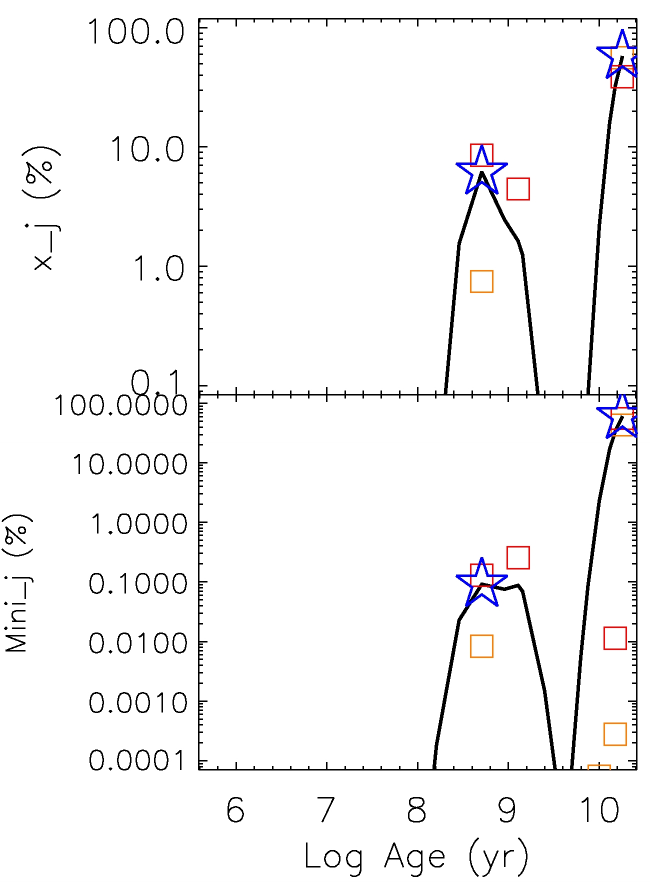}
  \caption{Age vs. mass fraction (bottom) and light fraction (top) of the stellar populations 
  synthesized in the fitting of the spectrum of the X-shaped source J1201-0703. The squares correspond to stellar populations with different metallicities (in the color version: orange 1.0 Z$_{\odot}$ and red 2.5 Z$_{\odot}$). The solid curve represents Gaussian smoothing of the mass fraction and light fraction distributions. The resulting peaks or bursts of star formation are marked with a star.}
\label{figure2}
\end{figure}

The STARLIGHT model for the observed spectra yields an estimate of
stellar velocity dispersion $\sigma_{*}$, which we corrected by taking into account the spectral resolution of each object, and from which we derived the BH mass using the empirical $M_\mathrm{BH} - \sigma_{*}$ relation
(\citealt{2000ApJ...539L..13G}; \citealt{2002ApJ...574..740T}), and
the light fraction, mass fraction, age, and metallicity of the stellar
populations used in the fit. We used these parameters to derive
starburst histories and apply Gaussian smoothing to the individual
starburst events in order to determine the epoch of the most recent
starburst episode (see Fig.~\ref{figure2}).  A quality factor $Q$ was
derived from the $\chi^{2}$ of the modeled spectra in order to
quantify the reliability of the fit. Fits with $Q > 10$ can be
considered reliable (see M11 for details).

We estimated the rest-frame continuum flux at 5100 \AA\ from the SDSS
photometry (\citealt{2004ApJ...614...91W}).  To assess the spectral
classification of the X-shaped radio galaxies, we measured the Ca II
break (C$_\mathrm{Ca\ II}$) of their absorption optical spectrum
(\citealt{2002MNRAS.336..945L}) as done in M11.

Finally, we determined the dynamic age ($t_{a}$) of the
high-surface-brightness (active) radio lobes using
$\theta_\mathrm{a}/v_\mathrm{a}$, where $\theta_\mathrm{a}$ is the
angular size in the FIRST image, converted to linear size assuming that the sources are in the plane of the sky, and $v_\mathrm{a}$ is the lobe advance
speed. We adopted $v_\mathrm{a} \approx 0.1\,c$, as obtained from observations of the aging of the synchrotron radiation spectrum in radio galaxies (e.g., \citealt{1995MNRAS.277..331S}; \citealt{1998AJ....115..960T}).

Assuming that the fueling of the low-surface-brightness lobes of the
X-shaped sources had stopped after the high-surface ones were
activated, the dynamic age of the passive lobes $t_{p}$ during their
active stage can be estimated as
\begin{equation}\label{equation1}
 t_{p}= \frac{\theta_{p}-t_{a}v_{p}}{v_{a}}\,,
\end{equation}
where $\theta_{p}$ is the angular size of the low-surface brightness
lobes, converted to linear size assuming that the sources are in the plane of the sky, and $v_{p}$ is their expansion speed during the inactive
stage. We use $v_{p}= 0.01\,c$ in our calculation.

\begin{figure*}[t!]
 \centering
\includegraphics*[width=0.8\textwidth]{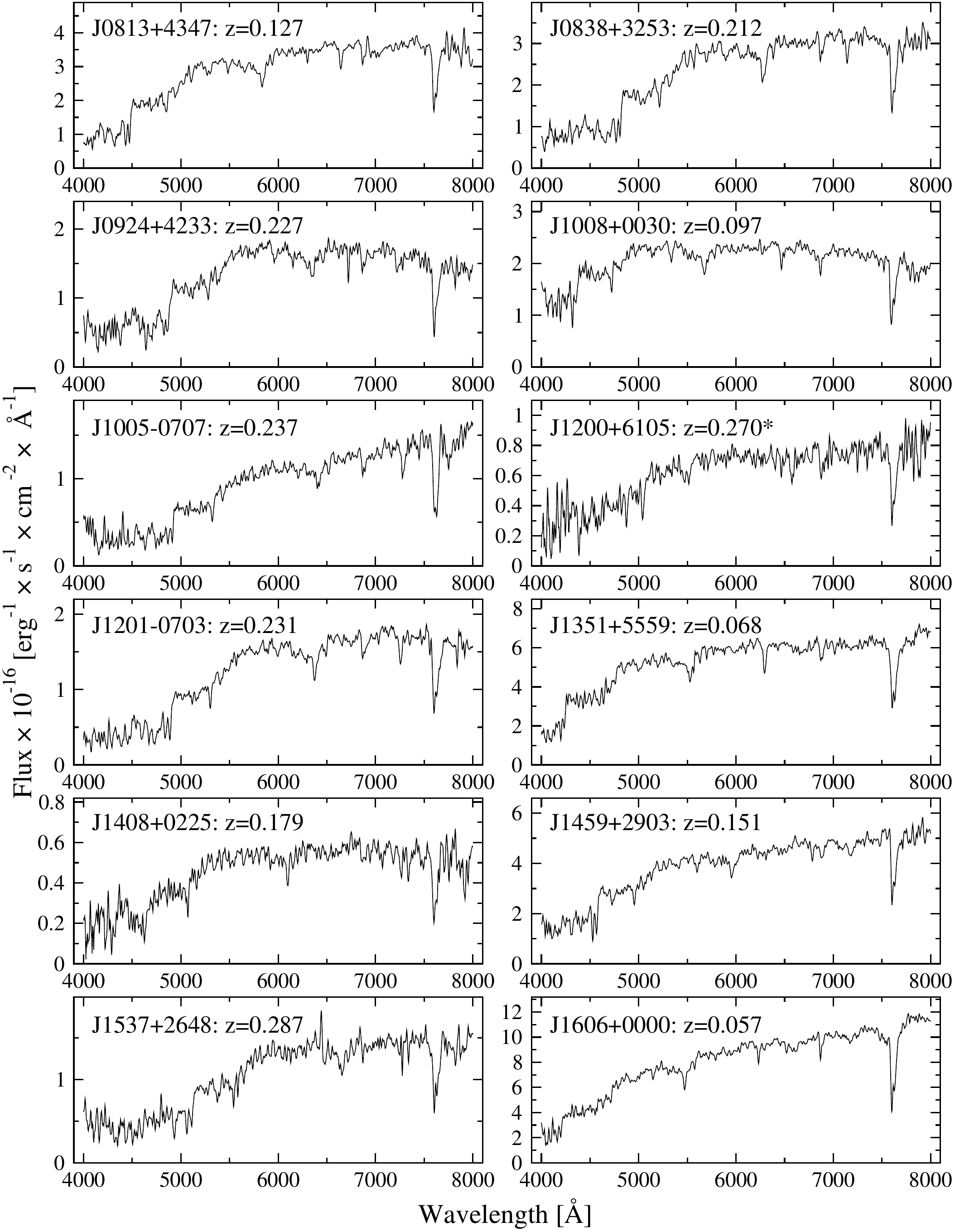}
  \caption{Optical spectrum and redshift of the targeted X-shaped radio sources. *The noisy spectrum of J1200+6105 results in a higher uncertainty of the redshift determination for this object.}
\label{figure3}
\end{figure*}

\section{Results}

\begin{figure}
 \centering
\includegraphics*[width=\columnwidth]{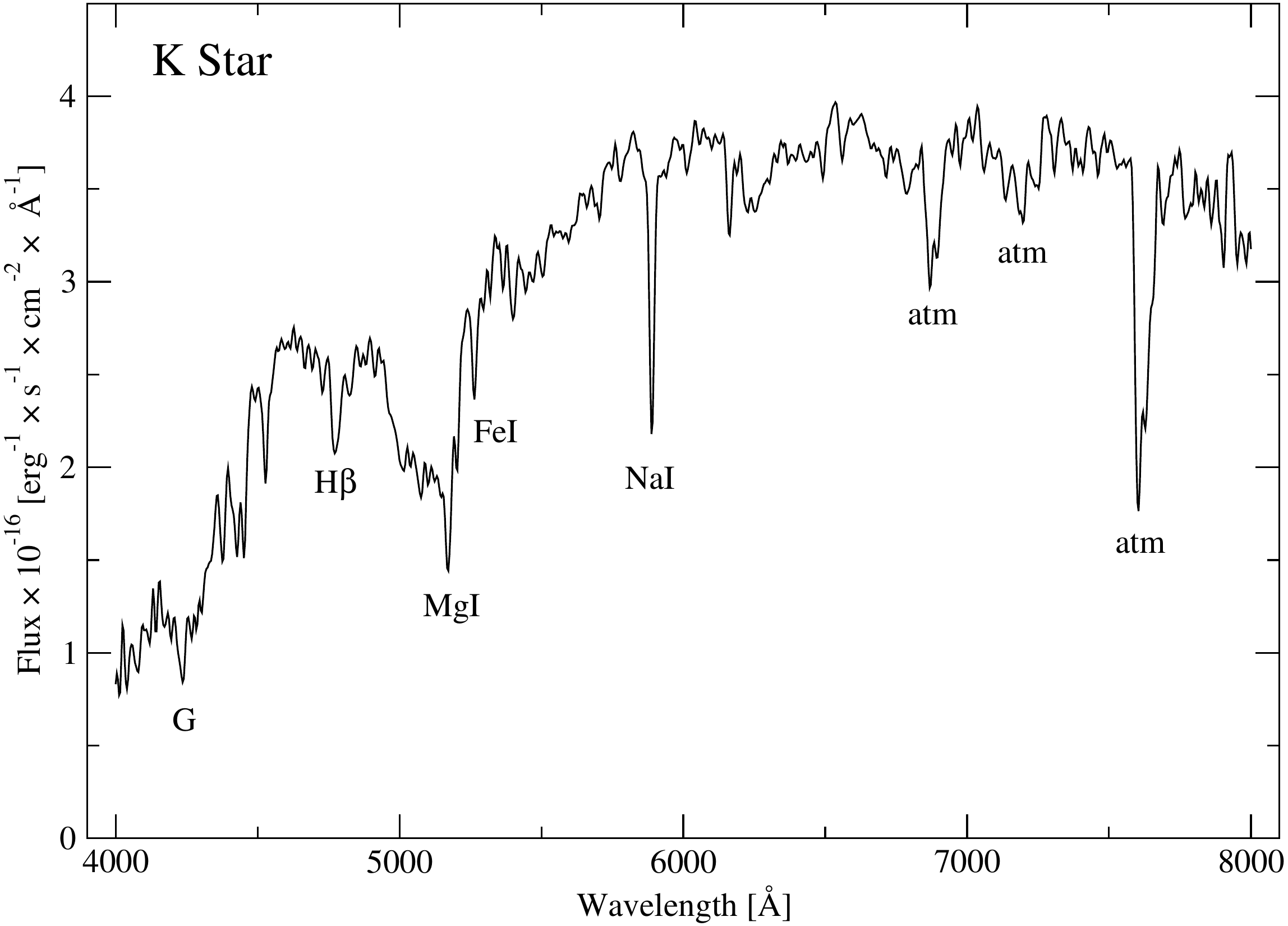}
  \caption{Spectrum of the optical counterpart of J1258+3227, showing stellar lines typical of a K-type star. This object is excluded from our analysis.}
\label{figure4}
\end{figure}

The new data provide the first spectroscopic identifications and
redshifts for 8 X-shaped radio galaxies (J1008+0030, J1055-0707,
J1200+6105, J1201-0703, J1351+5559, J1408+0225, J1459+2903, and
J1537+2648). New spectra were obtained for another 4 X-shaped sources
(J0813+4347, J0838+3253, and J0924+4233, spectroscopically identified
in the SDSS, \citealt{2007ApJS..172..634A}, and for J1606+0000,
identified by \citealt{1999MNRAS.310..223B}). The difference between
our redshifts and the ones obtained previously are $\delta_\mathrm{z}
< 0.002$. The spectrum of these 12 sources is shown in Fig.~\ref{figure3}. With the new identifications, the total number of X-shaped
radio galaxies spectroscopically identified is increased to 60. The
last object of our sample, J1258+3227, shows the typical stellar lines of
a K-type star (G band, H$\beta$, Mg I, Fe I, and Na I absorption
lines; Fig.~\ref{figure4}), indicating that there has been a likely misidentification (as also suggested by
\citealt{2009ApJS..181..548C}). This object is excluded from further
analysis. Combined results from the fits to the optical spectra, the
BH mass calculations, the epochs of the most recent starburst, and the
age estimates for the radio lobes are presented in Table~2. The
columns listed are: object name based on J2000.0 coordinates (Col. 1),
other common catalog names (Col. 2), stellar velocity dispersion
(Col. 3), BH mass derived from $\sigma_{*}$ (Col. 4), optical
luminosity of the AGN (Col. 5) and of the host galaxy (Col. 6), radio
luminosity (Col. 7), dynamic age of the radio lobes (Col. 8), most
recent starburst age (Col. 9), the value of Ca II break factor
(Col. 10), spectroscopic redshift (Col. 11), and quality factor of the
STARLIGHT fit (Col. 12). The total (active $+$ passive lobe) age of
the radio emission is given in brackets in Col. 8.  The objects
J0813+4347, J0838+3253, and J0924+4233 were previously analyzed in
M11. We report here their new values of stellar velocity dispersion,
BH mass, most recent starburst age, Ca II break factor,
spectroscopic redshift, and quality factor ($Q$) of the STARLIGHT fit
derived from the new spectra. The new fits have higher values of $Q$
than in previous analysis and are used for the combined
statistical study.

The addition of the new spectroscopical identifications to the
X-shaped sample of M11 provides an extended X-shaped sample of 38
sources. This extended sample is considered from now on for the
statistical comparison against a control sample consisting of 36
radio-loud sources with redshift z$<$0.3, and covering the same range of optical and radio luminosities as the X-shaped sample. A control subsample of elliptical radio-loud sources with values of $C_\mathrm{Ca\ II}>0.25$ and that qualify as early-type galaxies according to 
color-color diagnostics (i.e., location above the \textit{u}-\textit{r} galaxy type separator in a color-color diagram; \citealt{2001AJ....122.1861S}) is also considered for the statistical comparison (selection of both samples is described in detail in M11). The morphological classification based on visual inspection of SDSS images of this subsample of elliptical control sources agrees with the one based on the Ca II break and color-color diagram.

\subsection{Luminosity matching}

To provide a closer statistical matching between the target
and control samples, we used tight radio and optical luminosity
matching between the two samples.

\begin{figure}
 \centering
\includegraphics*[width=\columnwidth]{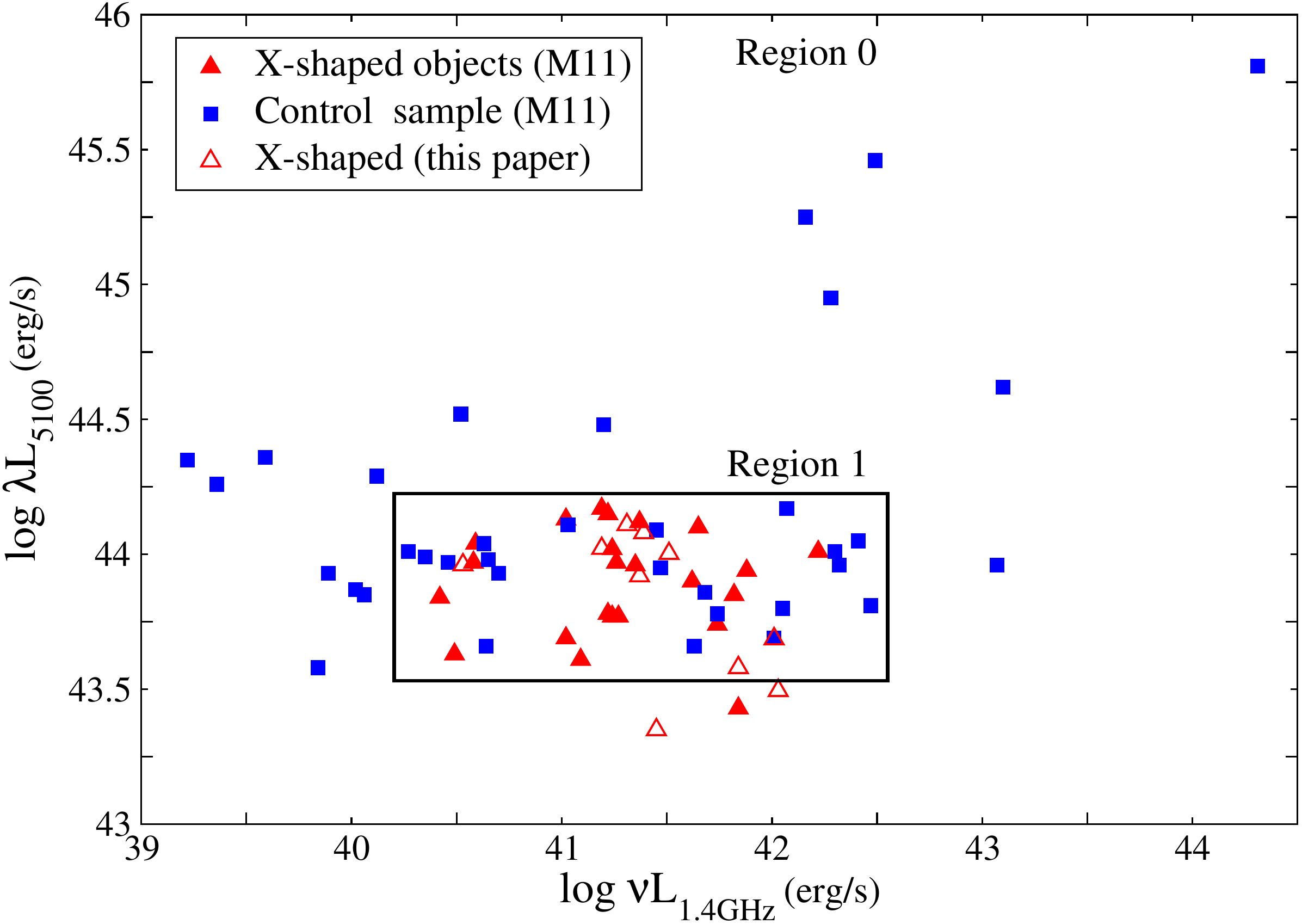}
  \caption{Optical continuum luminosity
  versus radio luminosity at 1.4 GHz for the X-shaped sources of M11 (filled triangles), control sources 
(squares), and the new X-shaped sources analyzed here (empty triangles). A small subregion (``Region 1") marked by a rectangle inside the plot is identified to provide a tighter luminosity match between the X-shaped and control samples (see M11 for details).}
\label{figure5}
\end{figure}

We derived the optical luminosities of all of the 12 X-shaped radio galaxies
except for two (J1055-0707 and J1201-0703), whose rest-frame continuum
flux at 5100 \AA\ cannot be estimated owing to their lack of SDSS
photometry. These sources are excluded from the statistical
study. Figure~\ref{figure5} compares the radio and optical
luminosities in the combined target and control samples (with the new
X-shaped objects marked).  All new sources fall in the common range of
luminosities $\log\,\lambda L_{5100 \AA} \in  [43.0,\,46.0]$ and $\log\,
\nu L_\mathrm{1.4GHz} \in  [39.0,\, 44.5]$, defined in M11 as Region
0, and eight of them fall within a tighter range of luminosities
called Region 1 (with $\log\,\lambda L_{5100 \AA} \in [43.5,\, 44.25]$ and
 $\log\,\nu L_{1.4GHz} \in [40.25,\, 42.5]$; see M11 for
further details of the region definition). 

The Kolmogorov-Smirnov (KS) test applied to the optical and radio
luminosities of the extended X-shaped sample and the control sample in
Region 0 gives a probability $> 5\%$ that the two samples
are drawn from the same parent distribution. 
In Region 1, the
KS-test indicates that the probability that the two samples are the
same is 87$\%$ for the optical luminosity and 19$\%$ for the
radio luminosity. Therefore, the KS tests suggest that the X-shaped and control
sample sources are indistinguishable in terms of optical and
radio luminosities within Region 1.

\subsection{Host type}

\begin{figure}
 \centering
\includegraphics*[width=\columnwidth]{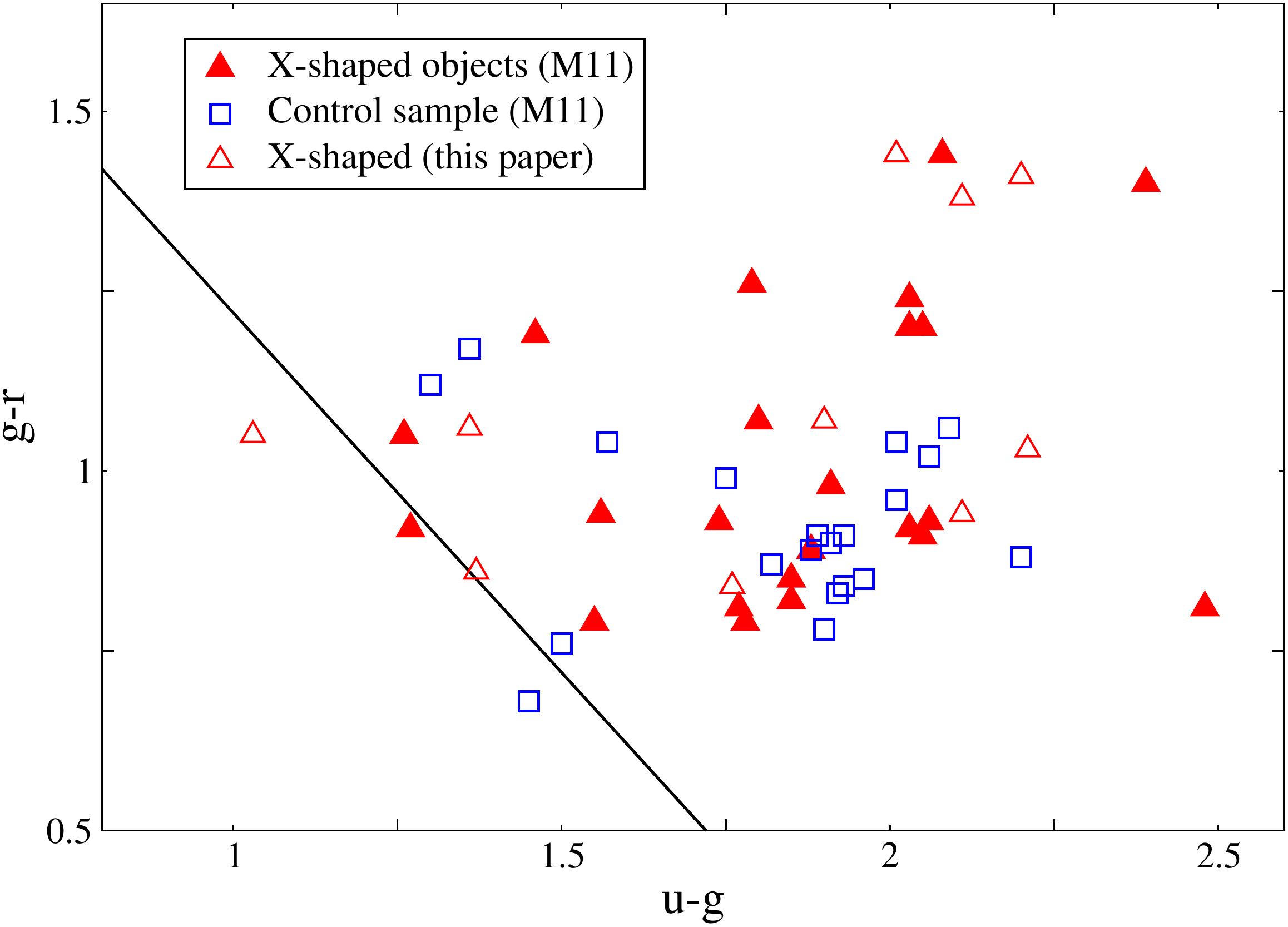}
  \caption{Color-color diagram ({\em g-r} colors versus {\em u-g} colors) for the 
  10 new X-shaped sources analyzed here (empty triangles), the X-shaped sources from M11 (filled triangles), and the control sources from M11 (squares).  Black line:
$u-r= 2.22$ galaxy type separator from \cite{2001AJ....122.1861S}. Sources situated above this line are classified as
early-type galaxies.}
\label{figure6}
\end{figure}

The new spectra present typical features of early-type galaxies,
showing strong metal absorption lines but no optical emission lines,
and with values of the Ca II break of C$_\mathrm{Ca II}$ $\geq$ 0.3 in
all the sources, and C$_\mathrm{Ca II}$ $\geq$ 0.4 in 10 of them (see
Table~2, Col. 10). These values ensure that the galaxy is dominated by
the thermal spectrum of the host rather than the non-thermal spectrum
of an active nucleus or a relativistic jet.  The host galaxy type can
be inferred using a color-color diagram, obtained from the SDSS $u$,
$g$, and $r$-band photometry. We apply a color-color distinction
(\citealt{2001AJ....122.1861S}) to identify the type of the host
galaxy in the 10 new X-shaped sources considered for the statistical
analysis. The combined plot presented in Fig.~\ref{figure6} shows that
the new sources also qualify as early-type galaxies.

\subsection{Black hole masses}

From the individual estimates of the BH mass, we calculated
the mean BH mass in the extended X-shaped sample, $\langle
M_\mathrm{BH,X-shaped}\rangle$, and derived its ratio to the mean BH
mass in the control samples $r_\mathrm{xc} = \langle
M_\mathrm{BH,X-shaped}\rangle/\langle
M_\mathrm{BH,control}\rangle$. The resulting ratio is
$1.49^{+0.24}_{-0.20}$ for the X-shaped/control samples and
$1.13^{+0.20}_{-0.17}$ for the X-shaped/control (ellipticals)
samples. These values increase to $1.93^{+0.42}_{-0.34}$ and
$1.50^{+0.36}_{-0.29}$, respectively, in the tighter Region 1. The
ratios obtained are similar to the ones obtained for the previously published
sample of M11, for which $r_\mathrm{xc}>1$ were found in
both Regions 0 and 1.
The median ratios are 
$1.03^{+0.29}_{-0.25}$ for the X-shaped/control samples and $1.01^{+0.25}_{-0.21}$ for the X-shaped/control (ellipticals) samples in Region 0, and
$1.03^{+0.43}_{-0.36}$ and $1.02^{+0.36}_{-0.30}$, respectively, in the tighter Region 1.
While the difference between the mean and the median values of the
mass ratios is significant, it should be noted that application of
the median is best justified for samples featuring a small number of
strong outliers, which is not the case for any of our samples.
Furthermore, the KS-test applied to
the BH mass distribution of the X-shaped and control sample gives
probabilities of 85$\%$ and 96$\%$ that the two samples differ
significantly in Regions 0 and 1, respectively, confirming the difference
between BH masses in the two samples expressed by the mean values of the mass.

\begin{figure}
 \centering
\includegraphics[width=\columnwidth, clip=true]{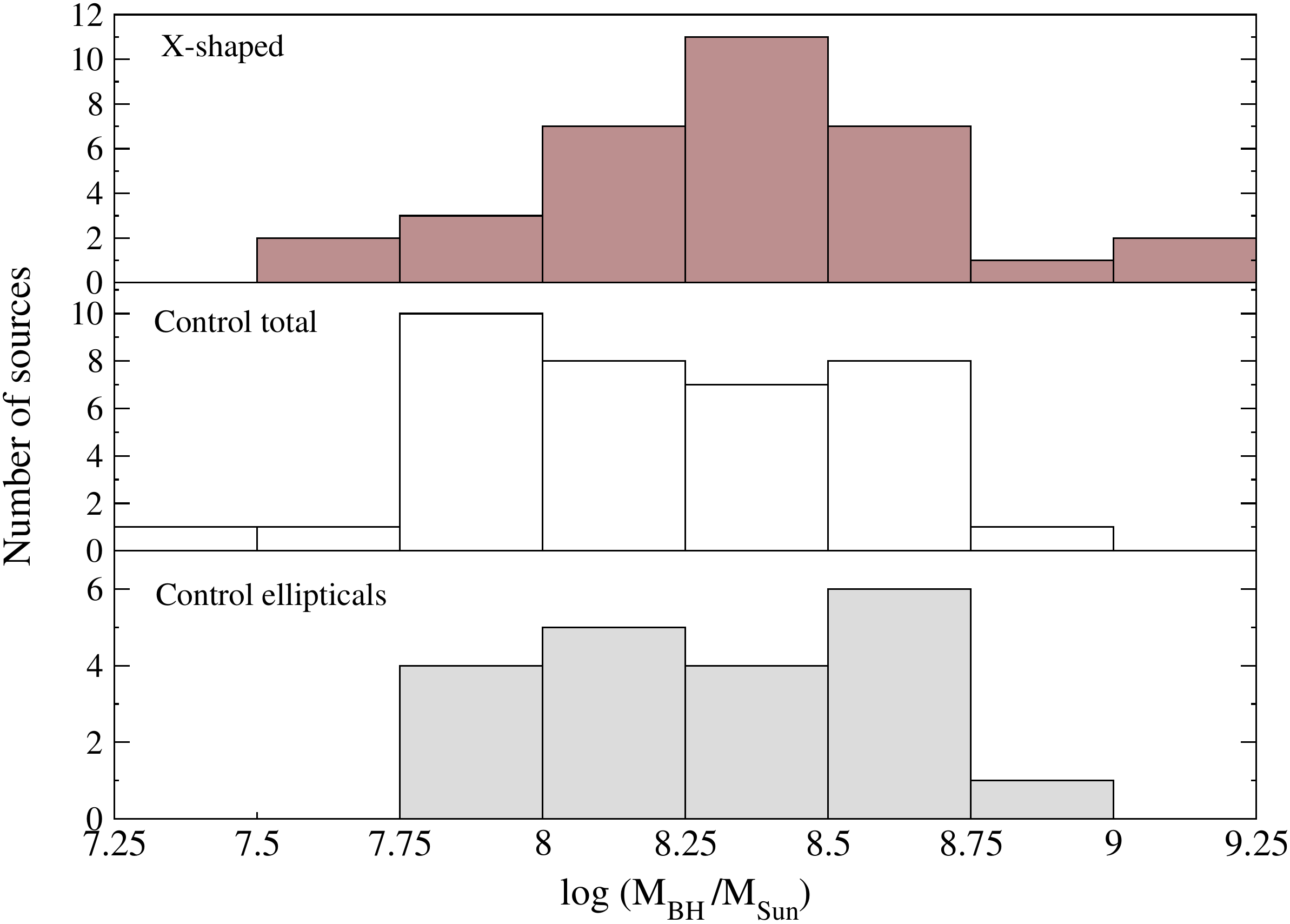}
  \caption{Histogram of the BH mass in Region 0 for the extended sample of X-shaped sources (top), entire sample of control galaxies (middle), and control ellipticals (bottom).}
\label{figure7}
\end{figure}

\begin{figure}
 \centering
\includegraphics[width=\columnwidth, clip=true]{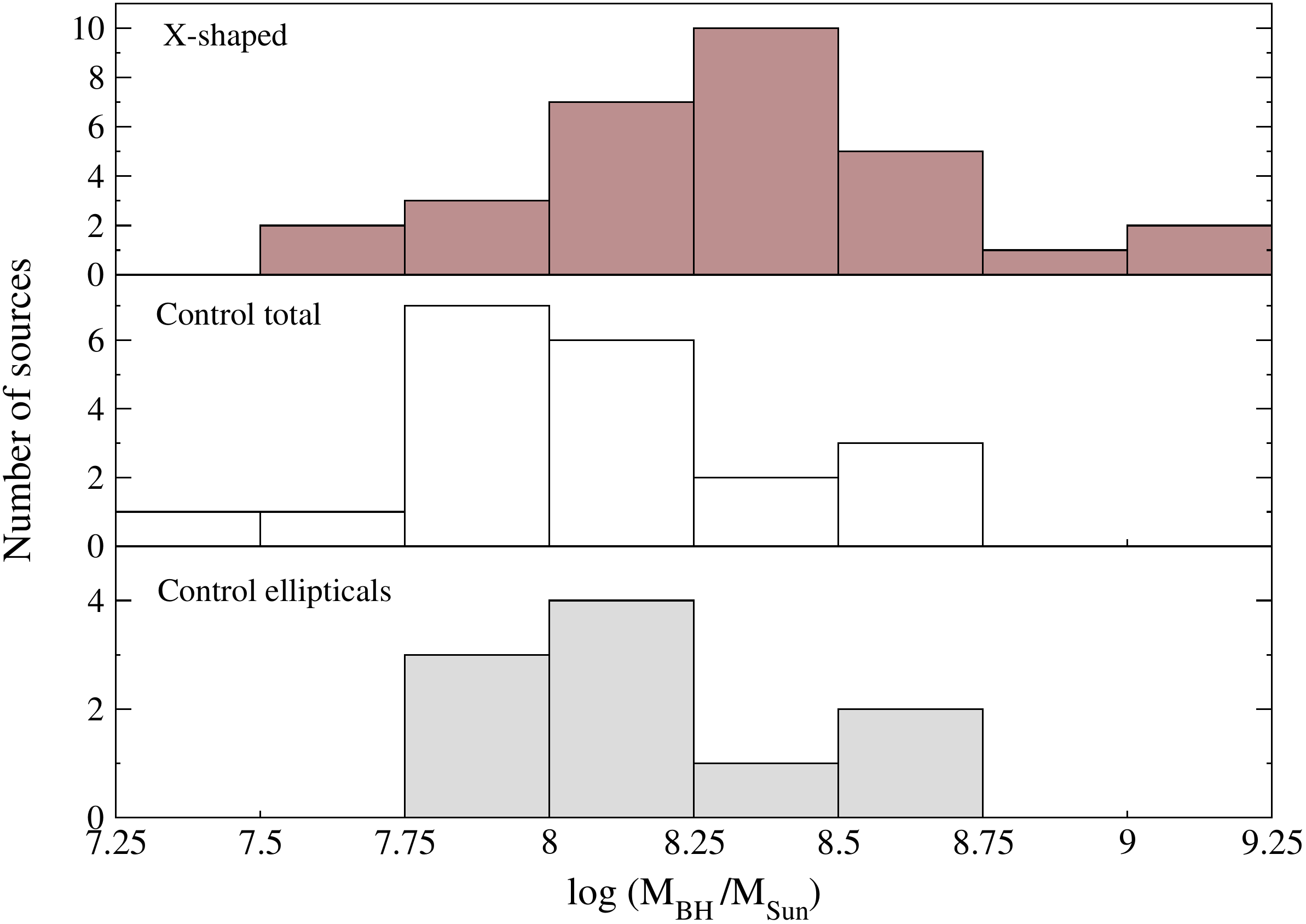}
  \caption{Histogram of the BH mass in Region 1 for the extended sample of X-shaped sources (top), entire sample of control galaxies (middle), and control ellipticals (bottom).}
\label{figure8}
\end{figure}

The differences between the BH masses of the two samples are
further illustrated by the distribution of BH masses in
Figs.~\ref{figure7}-~\ref{figure8}. The histograms of the mass
distributions for the X-shaped and control samples in Regions 0 and 1
underline the trend of the X-shaped objects having statistically
higher BH masses than the
control sample. About 85$\%$ of the X-shaped sources in Region 0 and 83$\%$ in Region 1 have
$\log$\,M$_\mathrm{BH} > 8$ M$_{\odot}$. The percentages for the control sample objects are 67$\%$ and 55$\%$ in Regions 0 and 1, respectively. These results lend further
support to the merger scenario in which a BH coalescence could
explain the higher BH masses obtained in the X-shaped sample and be
the possible origin of the X-shaped morphology.

\subsection{Starbursts}

\begin{figure}
 \centering \includegraphics[width=\columnwidth,
  clip=true]{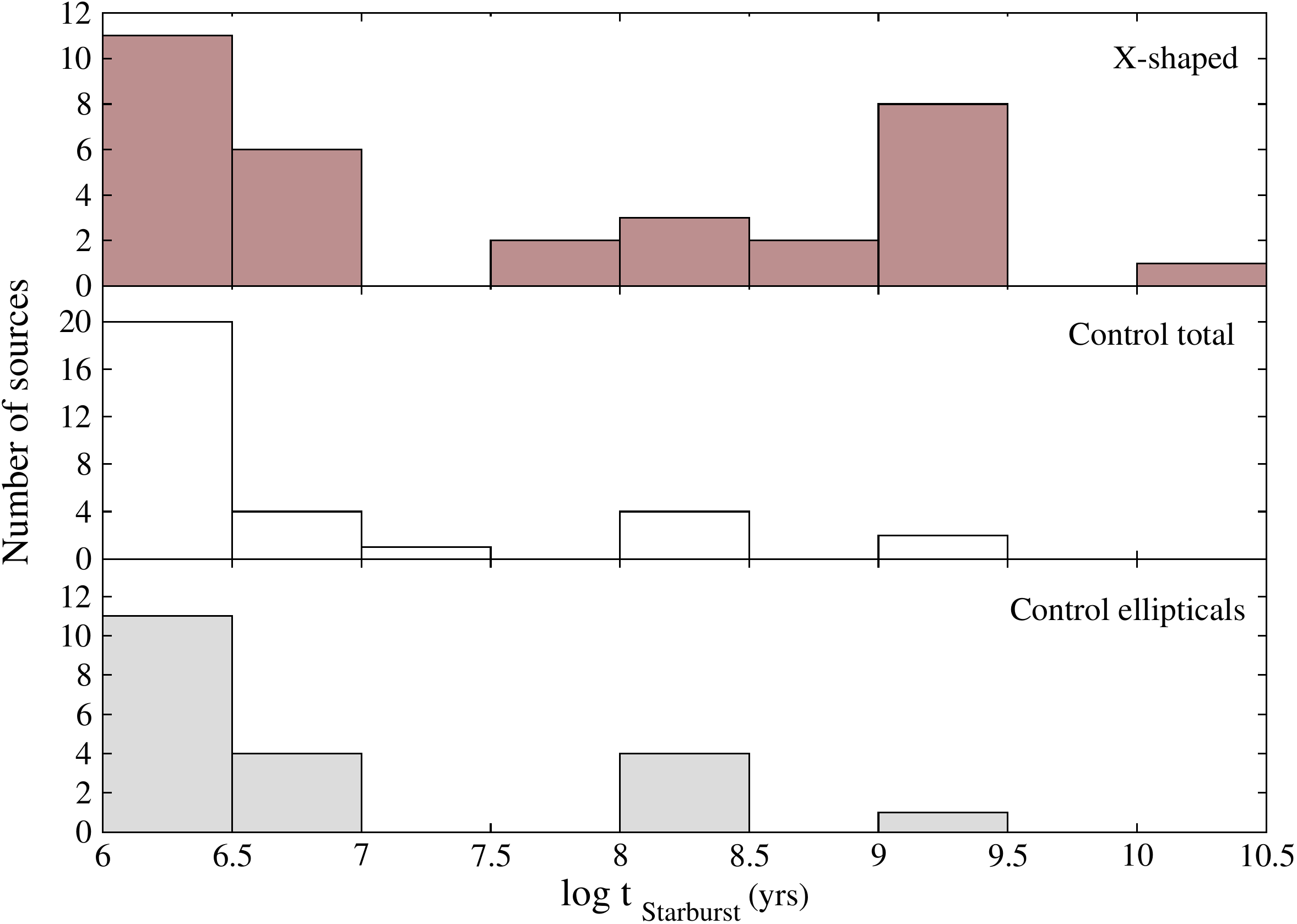} \caption{Histogram of
  the ages of the most recent starburst for X-shaped sources (top),
  entire sample of control galaxies (middle), and control ellipticals (bottom) in Region 0.}
\label{figure9}
\end{figure}

\begin{figure}
 \centering
  \includegraphics[width=\columnwidth, clip=true]{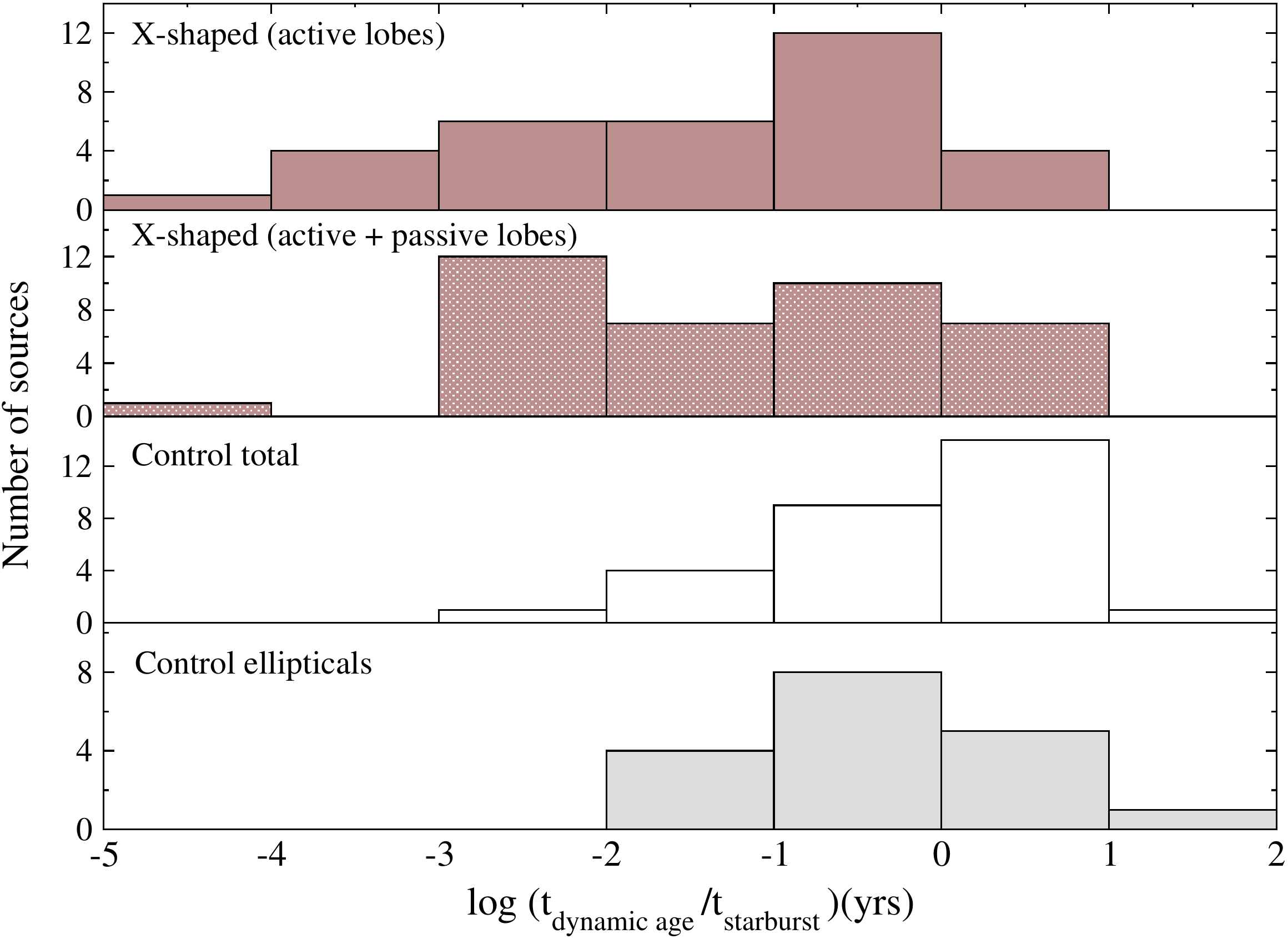}
  \caption{Logarithmic ratios of the dynamic ages of the radio emission to the
age of the most recent starburst for Region 0. The top row presents the ages of active
lobes in X-shaped objects, while the second row shows the sum of the
ages of the active and passive lobes. The two bottom rows show
distributions of the age ratios in the entire sample of control galaxies (third row) and
the control subsample of ellipticals (fourth row).}
\label{figure10}
\end{figure}

Distributions of the ages of the most recent starburst are compared in
Fig.~\ref{figure9} for the extended X-shaped sample and the control
samples. Both samples present a peak of starburst activity in the
range of $10^{6.0}$-$10^{6.5}$ years. The X-shaped sample notably exhibits
a more prominent spread in the starburst ages, with 50\% of
the sources having a starburst activity older than $10^{8}$ years. The
addition of the new galaxies to the X-shaped sample of M11 not only
confirms this trend but also reveals a secondary peak between
$10^{9.0}$-$10^{9.5}$ years in the distribution of starburst ages of
the extended X-shaped sample (see Fig.~\ref{figure9}).  The most
recent peak of star formation, which is present in all three samples, could be
related to the current jet activity (e.g.,
\citealt{1989MNRAS.239P...1R}; \citealt{2008A&A...477..491L};
\citealt{2009ApJ...700..262S}). The secondary peak, which is only present in
the extended X-shaped sample, cannot be linked to the active lobes as
their oldest dynamic age is $\sim10^{6}$ years (see
Table~\ref{table2}, col. 8). This strong starburst activity can only
possibly be related to an event that occurred before the active lobes
were formed.  A delay between a merger-driven starburst and AGN activity is shown in recent studies (e.g., \citealt{2010ApJ...714L.108S}; \citealt{2011MNRAS.412.2154B}). This scenario can be tested by
comparing the starburst ages to the dynamic ages of the radio lobes
for both the X-shaped and control samples. Histograms of the
logarithmic ratio of the dynamic age to the most recent starburst age
are shown in Fig.~\ref{figure10}.  The X-shaped sources tend to have
older starburst ages than the dynamic ages of the radio lobes, while
these two ages are comparable in the control sample. This trend in
the X-shaped sample is observed for both the active lobes alone and
the active plus passive lobes (see Fig.~\ref{figure10}). The mean
logarithmic ratio of the extended X-shaped sample is $-1.43 \pm 0.23$,
while that of the control sample is $-0.14 \pm 0.18$.  According to
the KS-test, the starburst ages of the X-shaped sample and the control
sample differ with a probability of $98\%$ in both Regions 0 and
1. Considering the subsample of only control ellipticals, the KS-test
gives a somewhat lower probability of $95.5\%$ of the
X-shaped objects being different from the control sample.

Since the dynamic ages have been derived from the linear sizes of the radio lobes, we check for differences in the linear size distributions of the X-shaped and control samples. The KS-test finds that the linear size distributions differ with a probability of 87\%. This difference is confirmed by the mean values of the distributions, which are (0.22 $\pm$ 0.02) Mpc for the X-shaped sample and (0.5 $\pm$ 0.2) Mpc for the control sample. The median values are 0.18 Mpc and 0.3 Mpc, respectively. We also study a possible dependence of the linear size with the BH mass and starburst age, but no dependences are found for any of the two samples.

\section{Discussion}

The origin of the X-shaped morphology in X-shaped radio galaxies is a
matter of ongoing debate.  Some recent studies in the X-rays
(\citealt{2010ApJ...717L..37H}) and optical
(\citealt{2011A&A...527A..38M}) bands and some theoretical models
(\citealt{2011ApJ...734L..32G}) favor a recent merger of two
supermassive black holes as the most plausible scenario. Other studies
(\citealt{2010ApJ...710.1205H}; \citealt{2010MNRAS.408.1103L}) propose a
backflow from the active lobes into the wings to explain the peculiar
morphology of X-shaped radio galaxies.

In this paper, we have presented the largest statistical study to date of the
BH mass and starburst activity in X-shaped radio sources. The
properties of a sample of 38 X-shaped radio galaxies have been
compared to those of a control sample of 36 radio-loud active nuclei,
by applying stringent luminosity and color criteria to provide the
closest match between the two samples. The results suggest that a
galactic merger is a more likely explanation of the origin of X-shaped
sources.  The merger hypothesis is further supported by the finding
that all of the X-shaped objects studied are hosted by early-type galaxies, which are expected to have undergone at least one major
merger event over the course of their evolution
(\citealt{1977egsp.conf..401T}).

Histograms of the BH mass distribution for the X-shaped and control
samples show that X-shaped radio galaxies tend to have higher
M$_\mathrm{BH}$ than the control sample. The ratio of the mean
M$_\mathrm{BH}$ for the two samples confirms the trend observed,
with the X-shaped sample having a mean BH mass that is $1.93^{+0.42}_{-0.34}$
times higher than the one of the control sample in a tight common range of
radio and optical luminosities. 
The higher average BH mass in
the galaxies hosting the X-shaped radio sources may result from a
major merger event and the consequent coalescence of the two
central black holes, which could be the origin of the winged
morphology observed in the radio maps. 

The occurrence of a merger event in the X-shaped radio galaxies should 
be reflected in their starburst histories. The age of the most recent burst of star
formation and the dynamic age of the radio lobes was thus studied for both
the X-shaped and the control samples. The most recent episodes of
starburst activity were found to be statistically older in the X-shaped sample
than those found for the control sample, with 50$\%$ of the
X-shaped sources having starburst ages older than 10$^{8}$ years. In
the X-shaped sources, the most recent starburst occurred before the
active lobes were formed. Of particular interest is the peak observed
in the distribution of the most recent starburst ages of the X-shaped
sample at ages of 10$^{9}$--10$^{9.5}$ years (1-3 Gyr). This peak
suggests that there has been enhanced star formation due to a merger event, which implies
a time delay of 1-3 Gyr between the peak of starburst activity and the
end of the merger. 
This timescale agrees with the time delay of $\sim$2 Gyr found by
hydrodynamical simulations of galaxy mergers
(\citealt{2008MNRAS.391.1137L}), and with the timescales of 0.5-2 Gyr found
observationally for individual objects (e.g.,
\citealt{2005MNRAS.356..480T}; \citealt{2006A&A...454..125E}). On the
other hand, the jet dynamic age of 10$^{6}$ yrs derived for the
X-shaped sources also agrees with the time delay between
the merger event and the onset of the radio-AGN triggered activity
found by other studies to be 10$^{6}$ years (e.g.,
\citealt{2006A&A...454..125E}).
These results assume a jet advance speed of 0.1\textit{c}. Estimates of the lobe advance speed based on synchrotron aging studies can reach 0.2\textit{c} for the most powerful radio galaxies, while the study of three samples of very powerful radio sources yielded estimates below 0.1\textit{c} (with a poor statistical significance; i.e., \citealt{1995MNRAS.277..331S}). The use of an advance speed as low as 0.01\textit{c} in our calculations would increase the dynamic age of the active lobes by an order of magnitude. However, this would not affect the results derived from the ratio of dynamic to starburst ages, which would still indicate that the most recent starburst activity occurred before the active lobe formation.
This result is also essentially unchanged when an advance speed of 0.2\textit{c} is used.

All the results obtained in our study strengthen the hypothesis that X-shaped radio
galaxies were formed by mergers. More detailed studies of star formation and gas kinematics
in these objects will yield stronger constraints of this scenario.



\section*{Acknowledgments}
The authors are grateful to the suggestions of the anonymous referee, which helped to improve the manuscript.
The authors thank M. Karouzos for insightful comments and discussions.
M. Mezcua was supported for this research through a stipend from the
International Max Planck Research School (IMPRS) for Astronomy and Astrophysics at the Universities of Bonn and Cologne.
This work was supported by the CONACYT research grant 54480 and 151494 (M\'exico).
The STARLIGHT project is supported by the Brazilian agencies CNPq, CAPES, and
FAPESP and by the France-Brazil CAPES/Cofecub program.

\begin{landscape}
\begin{table}
\small
\begin{minipage}[t]{\columnwidth}
\caption{\label{table2} X-shaped objects}
\centering
\begin{tabular}{cccccccccccc}
\hline
\hline
\multicolumn{2}{c}{Name} & $\sigma_\star$ & $\log\,M_\mathrm{BH}$ & $\log\,\lambda L_\mathrm{opt}^{agn}$ & $\log\,\lambda L_\mathrm{opt}^{gal}$ & $\log\,\nu L_\mathrm{rad}$ & $\log\,t_\mathrm{a}\,\,(\log\,t_\mathrm{p})$ & $\log\,t_\mathrm{sb}$ & $C_\mathrm{Ca\,II}$ & $z$ & $Q$ \\
J2000        & Other      & [km/s]          & [$M_\odot$]   & [erg/s]        & [erg/s]& [erg/s]& [yr]      & [yr] &          &       &      \\ 
(1)          & (2)        & (3)             & (4)           & (5)            & (6)   & (7)   & (8)         & (9)  & (10)     & (11)  & (12) \\ \hline
J0813$+$4347	&		&	274	$\pm$	83	&	8.7	$\pm$	0.5	&	$^{-a}$			&	44.11	&	41.31	&	6.29	\quad	(6.49)	&	9.11	&	0.45$^c$	&	0.127	&	259	\\
J0838$+$3253	&		&	237	$\pm$	79	&	8.4	$\pm$	0.6	&	$^{-a}$			&	43.92	&	41.37	&	6.59	\quad	(6.76)	&	8.71	&	0.50	&	0.212	&	88	\\
J0924$+$4233	&		&	266	$\pm$	102	&	8.6	$\pm$	0.7	&	$^{-a}$			&	43.58	&	41.84	&	6.54	\quad	(6.77)	&	9.11	&	0.42$^c$	&	0.227	&	117	\\
J1008$+$0030	&		&	340	$\pm$	41	&	9.1	$\pm$	0.2	&	41.73	$\pm$	1.37	&	44.02	&	41.19	&	5.61	\quad	(5.93)	&	6.00	&	0.32	&	0.097	&	709	\\
J1055$-$0707	&		&	261	$\pm$	46	&	8.6	$\pm$	0.3	&	$^{-a}$			&	$^{-b}$	&	42.24	&	6.14	\quad	(6.30)	&	7.40	&	0.47	&	0.237	&	72	\\
J1200$+$6105	&		&	165	$\pm$	74	&	7.8	$\pm$	0.8	&	43.27	$\pm$	0.20	&	43.69	&	42.01	&	6.08	\quad	(6.31)	&	6.70	&	0.29	&	0.270	&	115	\\
J1201$-$0703	&		&	303	$\pm$	43	&	8.9	$\pm$	0.2	&	$^{-a}$			&	$^{-b}$	&	41.71	&	5.94	\quad	(6.25)	&	8.71	&	0.50	&	0.231	&	149	\\
J1351$+$5559	&		&	228	$\pm$	78	&	8.4	$\pm$	0.6	&	$^{-a}$			&	43.96	&	40.53	&	5.42	\quad	(5.71)	&	7.00	&	0.49	&	0.068	&	116	\\
J1408$+$0225	&		&	266	$\pm$	53	&	8.6	$\pm$	0.3	&	41.53	$\pm$	1.24	&	43.35	&	41.45	&	5.67	\quad	(6.01)	&	10.26	&	0.39	&	0.179	&	313	\\
J1459$+$2903	&		&	153	$\pm$	77	&	7.7	$\pm$	0.9	&	41.97	$\pm$	4.57	&	44.00	&	41.51	&	5.98	\quad	(6.12)	&	6.94	&	0.38	&	0.151	&	142	\\
J1537$+$2648	&		&	219	$\pm$	42	&	8.3	$\pm$	0.3	&	$^{-a}$			&	43.49	&	42.03	&	6.05	\quad	(6.33)	&	9.11	&	0.40	&	0.287	&	33	\\
J1606$+$0000	&	4C$+$00.58	&	227	$\pm$	37	&	8.3	$\pm$	0.3	&	42.84	$\pm$	0.27	&	44.08	&	41.39	&	5.50	\quad	(5.71)	&	6.00	&	0.41$^c$	&	0.057	&	51	\\
\hline
\end{tabular}
\end{minipage}
\smallskip\newline {\bf Column designation:}~(1) -- object name based on J2000.0 coordinates; 
(2) -- other common catalog names; (3) -- stellar velocity dispersion obtained from STARLIGHT; 
(4) -- black hole mass obtained from $\sigma_{*}$; (5) -- 5100 \AA\ continuum
luminosity from STARLIGHT; (6) -- 5100 \AA\ continuum luminosity from
SDSS photometry; (7) -- 1.4 GHz radio luminosity; (8) -- dynamic age of the
active (active+passive) lobes; (9) -- age of the most recent starburst; (10) -- value of Ca II break factor; 
(11) -- spectroscopic redshift; (12) -- quality factor. {\bf Notes:}~$a$ -- STARLIGHT could 
not fit the continuum luminosity; $b$ -- no SDSS photometry available; $c$ -- values of Ca II break from \cite{2010MNRAS.408.1103L}.
\end{table}
\end{landscape}

\bibliographystyle{aa}
\bibliography{references}

\end{document}